\documentstyle[12pt,epsf, amstex]{article}
\newcommand{\be}{\begin{equation}}   
\newcommand{\ee}{\end{equation}}
\newcommand{\beqs}{\begin{eqnarray}}
\newcommand{\eeqs}{\end{eqnarray}}

\begin{document}
\pagestyle{plain}
\setcounter{page}{1}
\newcounter{bean}
\baselineskip16pt
%--------+---------+---------+---------+---------+--------

 \begin{titlepage}
 \begin{flushright}
SU...  \\
hep-th/9804120
\end{flushright}

\vspace{7 mm}

\begin{center}
{\huge Non commutative geometry and super \\ 
\vspace{3mm} 
Yang-Mills theory }

\end{center}
\vspace{12 mm}
\begin{center}
{\large
Daniela Bigatti } \\
\vspace{3mm}
Universit\`a degli Studi di Genova \\ 
Dipartimento di Fisica 
\vspace{10mm}

{\large Abstract} \\
\end{center}

\vspace{3mm}                                       

We aim to connect the non commutative geometry ``quotient space'' viewpoint 
with the standard super Yang Mills theory approach in the spirit of 
Connes-Douglas-Schwartz and Douglas-Hull description of application of 
noncommutative geometry to matrix theory. This will result in a 
relation between the parameters of a rational foliation of the torus and the 
dimension of the group $U(N)$. Namely, we will be provided with a 
prescription which allows to study a noncommutative geometry with rational 
parameter $p/N$ by means of a $U(N)$ gauge theory on a torus of size $\Sigma / 
N$ with the boundary conditions given by a system with $p$ units of magnetic 
flux. The transition to irrational parameter can be obtained by letting 
$N$ and $p$ tend to infinity with fixed ratio. The precise meaning of the 
limiting process will presumably allow better clarification.

\noindent

\vspace{7mm}
\begin{flushleft}
April 1998 

\end{flushleft}
\end{titlepage}
%--------+---------+---------+---------+---------+--------
\newpage
\renewcommand{\baselinestretch}{1.1}
% include the next line for double spacing

% \renewcommand{\baselinestretch}{2}

\section{Foliated torus and the noncommuting $U$, $V$ algebra}
We would like to clarify the relation between the so-called
non-commuting torus geometry and the picture of a torus foliation in the
situation where the slope is irrational (that is, the leaf is infinite).
Let us recall the main features of the two approaches.
\par In the noncommuting torus representation, one generalizes the toroidal
geometry by supposing that the two coordinates are not ordinary variables,
but satisfy the commutation relations
\begin{eqnarray} \displaystyle{
[p, q] = \frac{2 \pi i}{N}
} \end{eqnarray}
A more convenient labeling in order to provide an useful representation
for all $N \times N$ matrices is
\begin{eqnarray} \displaystyle{
U = e^{ip}
} \end{eqnarray}
\begin{eqnarray} \displaystyle{
V=e^{iq}
} \end{eqnarray}
so that
\begin{eqnarray} \label{commrel} \displaystyle{
UV = e^{2 \pi i / N} VU
} \end{eqnarray}
This gives a representation in the same spirit of the Fourier sum expansion:
\begin{eqnarray} \displaystyle{
Z= \sum_{n,m=1}^N {Z_{n,m} (U)^n (V)^m  }
} \end{eqnarray}
To any $N \times N$ matrix we thus associate a function of two periodic
variables.
\par Let's now switch to the foliated torus description. Let us take the
relation of equivalence which identifies, over the torus, the points belonging 
to the same straight line within a line of parallel straight line of fixed 
slope. The point is slightly tricky (since the torus itself is already born 
out of an equivalence relation : $\: T^2 \simeq {\Bbb R}^2 / {\Bbb Z}^2 \, $ ),
so let us be as explicit as possible. We consider the square unit torus 
(that is, $[0,1] \times [0,1]$ with the opposite sides ordinately glued). The 
correct procedure is to refer to the line bundle in ${\Bbb R}^2$ 
\begin{eqnarray} \displaystyle{
y=ax + b \: \: \: \: \: \: \: \: \: \: \: \: \: \: \: \:\: \: \: \: \: (a \: 
{\mathrm{fixed}})
} \end{eqnarray}
and to identify the points belonging to any such leaf. After this, we will 
proceed to carry on the equivalence which defines the torus. 
\par A leaf of the foliation is parametrized by the value of $b$, but in a 
very redundant way. Actually two values of the intercept correspond to the 
same leaf provided that 
\begin{eqnarray} \displaystyle{
b'= b+an \: \: \: \: \: \: \: \: \: \: \: \: \: \: \: \: \: \: \: \: \: 
n \in {\Bbb Z}
} \end{eqnarray}
Now let's introduce functions of two variables (one of which is an integer): 
$F(b, n)$. They actually admit to be interpreted as matrices (infinite, but 
with discrete entries), if one rewrites them in the form $F(b, b')$ where 
$b \sim b'$ (actually $b'= b+an$). For such functions we are about to define 
a multiplication structure, which, of course, will be inspired by the usual 
matrix product: 
\begin{eqnarray} \displaystyle{
H \equiv F * G 
} \end{eqnarray}
\begin{eqnarray} \label{c9} \displaystyle{
H(b, m) = \sum_{n} {F(b, n) \:  G(b+an, m-n)} 
} \end{eqnarray}
\par The matrices of (\ref{c9}) are actually parametrized by $b$, but if we 
replace $b$ with $b'= b+ ap$, this results, at the level of the matrix 
product, in a relabeling (by positions) of rows and columns of an infinite 
matrix. In other words, they are actually dependent on the leaf only. 
\par It is straightforward to check the matching of the $*$ definition and 
of the matrix product and, thus, the equivalence of the two descriptions. 
\par Switching to the rational case and choosing, moreover, $\theta = 1/N$, 
let us now give an explicit representation for the matrices $U$, $V$ 
which will turn out very useful in the following. Let's have as $U$ the 
shift matrix 
\begin{eqnarray}\label{10} \displaystyle{ 
U= \left( 
\begin{array}{cccccc}
0 & 1 & 0 & 0 & 0 & 0 \\
0 & 0 & 1 & 0 & 0 & 0 \\
0 & 0 & 0 & 1 & 0 & 0 \\
0 & 0 & 0 & 0 & 1 & 0 \\
0 & 0 & 0 & 0 & 0 & 1 \\
1 & 0 & 0 & 0 & 0 & 0 \\
\end{array}
\right) 
} \end{eqnarray} 
and as $V$ the matrix of phases 
\begin{eqnarray} \label{11} \displaystyle{ 
U= \left( 
\begin{array}{cccccc}
1 & 0 & 0 & 0 & 0 & 0 \\
0 & e^{2 \pi i / N} & 0 & 0 & 0 & 0 \\
0 & 0 & e^{4 \pi i / N} & 0 & 0 & 0 \\
0 & 0 & 0 & e^{6 \pi i / N} & 0 & 0 \\
0 & 0 & 0 & 0 & e^{8 \pi i / N} & 0 \\
0 & 0 & 0 & 0 & 0 & \ddots \\
\end{array}
\right) 
} \end{eqnarray} 
Both $U$, $V$ are to be interpreted as $N \times N$ matrices. 
\par It is straightforward to check that (\ref{10}), (\ref{11}) satisfy the 
commutation relation (\ref{commrel}). 

\section{Towards M-theory} 
We start the analysis by considering IIA theory on a torus and using Seiberg's 
view of matrix theory \cite{S}, that is, we replace the lightlike 
compactification with a compactification along a spacelike circle of shrinking 
radius. The paper \cite{DH} discusses a weakly coupled IIA theory as above with a 
background 2-form potential; let its value be $\theta$. We take a square 
torus, which gives immediately $\tau = i$. It will be convenient (for T 
duality purposes) to choose as parameters of the system 
\begin{itemize}
\item{$\tau \equiv i$}
\item{$P := \theta + ia$}
\end{itemize}
where $a$ is the area of the torus. 
\par Let's do now a T-duality transformation along one cycle of the torus, 
let's say, the ``1'' 1-cycle. Such a transformation interchanges $\tau$ and 
$P$; that is, we obtain a new torus with parameters 
\begin{eqnarray} \displaystyle{
P'= \tau
} \end{eqnarray}
\begin{eqnarray} \displaystyle{
\tau' = P 
} \end{eqnarray}
that is 
\begin{eqnarray} \displaystyle{
P'=i \:\:\:\:\:\:\:\:\:\:\: \Rightarrow \:\:\:\:\:\:\:\:\:\: a'=1 \: , 
\:\:\:\:\: \theta' = 0 
} \end{eqnarray}
\begin{eqnarray} \displaystyle{
\tau' = \theta +ia
} \end{eqnarray}
\par If the area gets small, the dual torus becomes very elongated (a sharp 
angle and the other nearly flat). 
\par What happens to a D0-brane under this duality? We will obtain a D1-brane 
oriented along the ``1'' direction of the new torus. (See figure 1.) Now 
suppose we have a fundamental string whose ends are on the D1-brane but which 
is wound $n$ times along the ``2'' 1-cycle. If we try to impose a constraint 
of minimal length (provided the winding number $n$ is fixed) and if we imagine 
to ``open up'' the torus and refer to a tiling of the plane made by its 
copies (see figure 2), we realize that the two points where the fundamental 
string is attached to the D-string are separated by a distance $n \theta$ 
along a straight line almost perpendicular to the D-string. In other words, 
they are equivalent points in the sense of the foliated torus construction 
described before. 
\par If we consider, at this stage, the field operators which create and 
annihilate such strings, they are fields whose arguments are two points 
belonging to the D-string and related by the equivalence relation of the 
foliated torus. In other words, it is now natural to interpret them as 
objects which live in the noncommutative algebra of the foliated torus.

\section{The rational foliation case}
\par Let us follow in this section the approach of \cite{DH} outlined in the 
previous pages, but with the additional assumption that the value $\theta$ 
of the background is rational: 
\begin{eqnarray} \label{cqq} \displaystyle{ 
\theta \equiv \frac{p}{N} \: \: \: \: \: \: \: \: \: \: \:  
p, \: N \: {\mathrm {relatively \: prime}}
} \end{eqnarray} 
The purpose of the present work is to connect the parameters of a super 
Yang-Mills theory on a torus (in particular the value of $N$) with the 
geometrical viewpoint of the noncommutative foliation. The first step 
will be to just assume $\theta = 1/N$. We will later handle the general 
case of (\ref{cqq}).         
\par Let us fix once more the notations (see figure 3). 
\begin{eqnarray} \displaystyle{ 
\begin{array}{ccc}
{ \begin{array}{l}
\left \{
\begin{array}{l}
\tau = i  \\
P = \theta + i \Sigma h 
\end{array}
\right .  
\end{array}
} & 
\stackrel{ {\mathrm {T \: duality } } }{ \longleftrightarrow }
& { \begin{array}{l}
\left \{
\begin{array}{l}
\tau' = \theta + i \Sigma h \\
P' =  i  
\end{array}
\right . 
\end{array} }
\end{array}
} \end{eqnarray} 
Let the horizontal axis be called $\sigma$, $0 \le \sigma \le \Sigma$. 
\par We have already seen that if we have a single D0-brane and make a 
T-duality transformation in the $\sigma$ direction, we obtain a single 
D1-brane oriented along the $\sigma$ direction. Moreover, strings compelled 
by ``minimal length constraints'' dynamically connect points $\sigma$ and  
$\displaystyle{\sigma + w \frac{\Sigma}{N}}$ where $w \in {\Bbb Z}$, thus 
defining (nonlocal) fields 
\begin{eqnarray} \label{-1} \displaystyle{ 
\phi(\sigma, w) 
} \end{eqnarray} 
To symmetrize the aspect of the above field, one might Fourier transform 
with respect to the periodic coordinate $\sigma$, thus obtaining a function 
of two integers: 
\begin{eqnarray} \label{0} \displaystyle{ 
\tilde{\phi} (k, w) \: \: \: \: \: \: \: \: \: \: \: k, \, w \in {\Bbb Z}
} \end{eqnarray} 
The energy of such a mode is 
\begin{eqnarray} \label{1} \displaystyle{ 
E(k, w) = \left( \frac{1}{\Sigma^2} \left[ k^2 + w^2 \right] 
\right)^{\frac{1}{2}} 
} \end{eqnarray} 
Let us now imagine of ``folding'' the circle according to the equivalence 
relation induced by the foliation (see figure 4). 
The small circle has now length $\displaystyle{\frac{\Sigma}{N}}$ and its 
periodic coordinate $x$ satisfies $\displaystyle{0 \le x \le \frac{\Sigma}{N} 
}$\,. 
\par We should also notice that the choice of $p$ in the numerator only 
corresponds to a relabeling of the $N$ sectors, which are rearranged in a 
permuted order. (See figure 5). 
This is a consequence of the (trivial) fact that, given $p$, $N$ 
relatively prime, no one of the numbers 
\begin{eqnarray} \nonumber \displaystyle{ 
p, \: 2p, \:, 3p, \: ... \: (N-1)p 
} \end{eqnarray} 
is a multiple of $N$ and thus 
\begin{eqnarray} \nonumber \displaystyle{ 
\{ 0, \: (p)_{{\mathrm{mod}}\: N}, \: (2p)_{{\mathrm{mod}}\: N} \:, ... \} 
} \end{eqnarray} 
is a set of $N$ integer numbers belonging to $[0, N]$ and all different. 
Thus choosing $p/N$ instead of $1/N$ results only in a relabeling of the 
equivalence relation: the ``arches'' are superimposed in a different order. 
%\par We want now to split the energy modes among the contribution of the 
%``true winding modes'' which are periodic of period $\Sigma$ and the 
%``fractional winding modes'' which are made possible by the equivalence 
%relation: 
\par We are now going to split the winding and the momentum modes into a 
part which is an integer multiple of $N$ and the remainder (see figure 6). 
%To this aim 
First we define an appropriate splitting of the integer $k$ of 
equation (\ref{0}): 
\begin{eqnarray} \label{*} \displaystyle{ 
k= KN + q \: \: \: \: \: \: \: \: \: \: \: \: \: \: \: \: K, q \in {\Bbb Z } 
\: \: \: \: \: \: \: \: 0 \le q < N
} \end{eqnarray} 
\par We simultaneously replace the fields $\phi$ of eq.~(\ref{-1}) with 
matrix valued fields 
\begin{eqnarray} \label{A} \displaystyle{ 
\phi_{ab} (x, W) \: \: \: \: \: \: \: \: \: \: \: a, b = 1, ...N 
} \end{eqnarray} 
where the index $a$ (resp.~$b$) tells us in which of the $N$ intervals 
(of length $\displaystyle{\frac{\Sigma}{N}}$ each) the string begins 
(resp.~ends) and the integer $W$ is the winding number along the big circle 
of length $\Sigma$. 
\par The relation with $w$ is given by 
\begin{eqnarray} \displaystyle{ 
w= NW + (a-b) 
} \end{eqnarray} 
We can now rewrite the energy (\ref{1}) in terms of the capital variables: 
\begin{eqnarray} \label{spectrum} \displaystyle{ 
E^2 = \frac{1}{\Sigma^2} \left( (NK+q)^2 + (NW+(a-b))^2  \right) 
} \end{eqnarray} 
\par We now wish to reproduce the energy spectrum (\ref{spectrum}) with a 
description on a small torus of size $\Sigma / N$. On such torus we will 
have two directions, $x$ and $y$ respectively. The description (\ref{A}) of 
the matrix fields can yield a description in one more direction if, as 
usual, we wish to interpret the winding mode as a Kaluza-Klein momentum in 
an additional direction; that is, Fourier transformation with respect to 
$W$ gives 
\begin{eqnarray} \displaystyle{ 
\phi_{ab} (x, y) 
} \end{eqnarray} 
Of course, one might also Fourier transform (\ref{A}) with respect to $x$: 
\begin{eqnarray} \displaystyle{ 
\tilde{\phi}_{ab} (K, W) 
} \end{eqnarray} 
\par In order to reproduce the spectrum (\ref{spectrum}) we will introduce a 
background $U(N)$ Yang-Mills field. We will also have to introduce non 
trivial boundary conditions for the fields when trasporting them around 
the $x$ axis. The first couple of condition we impose are 
\begin{eqnarray} \displaystyle{ 
\phi_{a, b} (x + \frac{\Sigma}{N}, y)  = \phi_{a+1, b+1} (x, y)
} \end{eqnarray} 
\begin{eqnarray} \displaystyle{ 
\phi_{ab} (x, y+ \frac{\Sigma}{N}) = \phi_{ab} (x, y)
} \end{eqnarray} 
% {\Large what about $p/N$ ??? \\ }
that is, if we move along $x$ direction making a complete turn on the small 
circle we get a unit shift of both indexes (of the ``begin'' and ``end''  
sectors on the big circle), while the $y$ direction is associated to the 
``big'' winding number. In matrix language  
\begin{eqnarray} \displaystyle{ 
\phi(x + \frac{\Sigma}{N}, y)  = U^\dagger \phi(x, y) U 
} \end{eqnarray} 
\begin{eqnarray} \displaystyle{ 
\phi(x, y+ \frac{\Sigma}{N}) = \phi(x, y)
} \end{eqnarray} 
where $U$ is the $N \times N$ shift matrix 
\begin{eqnarray} \displaystyle{ 
U= \left( 
\begin{array}{cccccc}
0 & 1 & 0 & 0 & 0 & 0 \\
0 & 0 & 1 & 0 & 0 & 0 \\
0 & 0 & 0 & 1 & 0 & 0 \\
0 & 0 & 0 & 0 & 1 & 0 \\
0 & 0 & 0 & 0 & 0 & 1 \\
1 & 0 & 0 & 0 & 0 & 0 \\
\end{array}
\right) 
} \end{eqnarray} 
Along the $y$ direction we wish to introduce Wilson loops, in order to mimic 
the fractional contribution to the momentum which we encountered in 
eq.~(\ref{spectrum}): 
\begin{eqnarray} \displaystyle{ 
W(x) = exp \left( i \oint A_y (x,y) \: dy  \right)
} \end{eqnarray} 
We assume $A_x =0$ and $A_y$ independent of $y$: 
\begin{eqnarray} \label{c1} \displaystyle{ 
W(x) = exp \left( i \frac{\Sigma}{N} A_y(x)  \right) 
} \end{eqnarray} 
We put in a background vector potential: 
\begin{eqnarray} \label{c2} \displaystyle{ 
W (x) = exp \left( i \frac{x}{\Sigma}   \right) V 
} \end{eqnarray} 
where the matrix $V$ is 
\begin{eqnarray} \displaystyle{ 
V= \left( 
\begin{array}{cccccc}
\ddots & 0 & 0 & 0 & 0 & 0 \\
0 & e^{- 2 \pi i / N } & 0 & 0 & 0 & 0 \\
0 & 0 & 1 & 0 & 0 & 0 \\
0 & 0 & 0 & e^{ 2 \pi i / N } & 0 & 0 \\
0 & 0 & 0 & 0 & e^{4 \pi i / N } & 0 \\
0 & 0 & 0 & 0 & 0 & \ddots \\
\end{array}
\right) 
} \end{eqnarray} 
Notice that the background satisfies the same conditions of $\phi$ if we  
make a complete turn on the small circle: 
\begin{eqnarray} \displaystyle{ 
\Gamma(x + \frac{\Sigma}{N}) = U^{\dagger} \Gamma (x) U 
} \end{eqnarray} 
\par From (\ref{c1}) and (\ref{c2}) it follows immediately for the vector 
potential 
\begin{eqnarray} \displaystyle{ 
A_y = \frac{x}{\Sigma^2} N \: I + \frac{1}{\Sigma} 
\left( 
\begin{array}{cccccc}
\ddots & 0 & 0 & 0 & 0 & 0 \\
0 & -2 & 0 & 0 & 0 & 0 \\
0 & 0 & -1 & 0 & 0 & 0 \\
0 & 0 & 0 & 0 & 0 & 0 \\
0 & 0 & 0 & 0 & 1 & 0 \\
0 & 0 & 0 & 0 & 0 & \ddots \\
\end{array}
\right) 
} \end{eqnarray} 
where $I$ is the $N \times N$ identity matrix. 
\par The vector potential can thus be split in an ``abelian'' and in a  
``non abelian'' part. One should notice how the form of the abelian gauge 
field corresponds to a unit of abelian magnetic flux through the torus. 
\par Now let's consider the free part of the action for 
the scalars 
\begin{eqnarray} \displaystyle{ 
{\cal L} = \int \: dx \: dy \: Tr (\dot{\phi}^2 - (\nabla \phi)^2 )
} \end{eqnarray} 
where the covariant derivative is defined as usual 
\begin{eqnarray} \displaystyle{ 
\nabla \phi = \partial \phi + i [A, \phi] 
} \end{eqnarray} 
Let us consider separately the spectra of $\nabla_x$ and $\nabla_y$. 
\par First of all, the spectrum of $\nabla_x \equiv \partial_x$ has to be 
evaluated on a circle of length $\displaystyle{\frac{\Sigma}{N}}$. But 
actually we imposed a condition which allows $\phi$ to return to its value 
only after $N$ turns, so the system behaves as the circle was effectively 
of size $\Sigma$. The spectrum will be then 
\begin{eqnarray} \displaystyle{ 
\frac{i}{\Sigma} [KN + p] \: \: \: \: \: \: \: \: \: \: \: p \in {\Bbb{Z}} 
\, , \: \: 0 \le p \le N-1  
} \end{eqnarray} 
with the same decomposition of eq.~(\ref{*}).   
\par Second, let's turn our attention to $\displaystyle{\nabla_y =  
\partial_y + i [A_y, \cdot ]}$\,. The derivative piece has the usual 
spectrum on the circle 
\begin{eqnarray} \displaystyle{ 
i \frac{NW}{\Sigma} \: \: \: \: \: \: \: \: \: \: \: W \in {\Bbb{Z}}
} \end{eqnarray} 
since $\displaystyle{\phi(y + \frac{N}{\Sigma})= \phi(y)}$. The commutator  
(for which the ``abelian'' part of the vector potential can be dropped) is 
rewritten 
\begin{eqnarray} \displaystyle{ 
[A_y, \phi]_{a,b} = \frac{1}{\Sigma} (a-b) \phi_{a,b}
} \end{eqnarray} 
and the spectrum of $\nabla_y$ is 
\begin{eqnarray} \displaystyle{ 
\frac{i}{\Sigma} \left[ NW + (a-b) \right] 
} \end{eqnarray} 
%%%%%%%%%%%%%%%% {\Large There is a $2\pi$ error \\} maybe there are still... 
Since the equation of motion of $\phi$ is 
\begin{eqnarray} \displaystyle{ 
\ddot{\phi}  = ( \nabla_x^2 + \nabla_y^2 ) \phi 
} \end{eqnarray} 
the spectrum will be 
\begin{eqnarray} \displaystyle{ 
E^2 = \left[  \left( \frac{KN + p}{\Sigma} \right)^2 + \left( 
\frac{WN + (a-b)}{\Sigma} \right)^2  \right] 
} \end{eqnarray} 
as required by eq.~(\ref{spectrum}).

\section{Interaction terms}
To isolate our main point, we will refer to the quartic scalar interaction 
term. What happens is of course more general and could actually be imagined 
from the remarks in section 3; nevertheless, we want to proceed to an 
explicit exhibition of how the star product works as a substitute of matrix 
multiplication and to discuss the role of the numerator of the fraction 
$\theta \equiv p/N$. 
\par In the 2+1 dimensional Connes-Douglas-Schwartz theory over the 
``big'' torus, the form of the quartic terms is 
\begin{eqnarray} \label{quartic} \displaystyle{ 
\int_0^\Sigma {dX \: dY \: [\phi^i * \phi^j - \phi^j * \phi^i  ]^2 }
}\end{eqnarray} 
where $X$, $Y$ are coordinates on the ``large'' torus and the star product 
is defined as  
\begin{eqnarray} \displaystyle{ 
F * G = F(x,y) \: e^{i \frac{\theta}{2} (    
{ \stackrel{\leftarrow}{\partial}}_x \vec{\partial}_y 
- { \stackrel{\leftarrow}{\partial}}_y \vec{\partial}_x  )} \: G(x,y) 
} \end{eqnarray} 
For the moment we will assume $\theta \equiv 1/N$. 
\par To evaluate this, we go to the Douglas-Hull representation by Fourier 
transforming with respect to $y$: 
\begin{eqnarray} \displaystyle{ 
F * G = \sum_{n, m} F(x, n) \: e^{i n y} \: e^{i \frac{\theta}{2} (    
{ \stackrel{\leftarrow}{\partial}}_x \vec{\partial}_y        
- { \stackrel{\leftarrow}{\partial}}_y \vec{\partial}_x  )}  
\: \tilde{G}(x, m) \: e^{imy}
} \end{eqnarray} 
We replace ${ \stackrel{\leftarrow}{\partial}}_y$ 
by $in$ and ${ \stackrel{\rightarrow}{\partial}}_y$ by $im$: 
\begin{eqnarray} \nonumber \displaystyle{ 
F * G = \sum_{n, m} {F(x,n) \: e^{-m \theta /2 \vec{\partial}_x }   \: 
e^{n \theta / 2 \vec{\partial}_x }  \: G(x,n) \: e^{i(n+m)y } } = 
} \end{eqnarray} 
\begin{eqnarray} \displaystyle{ 
= \sum F(x- \frac{m \theta}{2}, n) \: G(x+ \frac{n \theta}{2}, m) 
\: e^{i(n+m)y} 
} \end{eqnarray} 
\par We now recall the intuitive picture of $F(x,n)$ as a ``string''
attached at $x$ and with length $n \theta$ along the $x$ axis (see figure 7). 
If we regard the strings on a given leaf (that is, the ones whose 
endpoints belong to that leaf, i.~e.~are equivalent with respect to the 
leaf induced relation of equivalence) as matrices (since the endpoints 
of $F$ and $G$ coincide), then the Fourier transform of $F*G$ is 
exactly the matrix product: 
\begin{eqnarray} \displaystyle{ 
\sum_n F(x- \frac{k-n}{2} \theta, n) \: G(x + \frac{n \theta}{2}, (k-n))   
} \end{eqnarray} 
(remember $m=k-n$). 
\par Notice that a given leaf corresponds to a point on a small circle. 
It was actually to the purpose of removing (this particular kind of) 
non locality that the small circle has been introduced. 
\par Similarly, expressions like 
\begin{eqnarray} \displaystyle{ 
\int (F*G)(H*K) \: dx \: dy 
} \end{eqnarray} 
translate in matrix language 
\begin{eqnarray} \displaystyle{ 
\int_0^{\frac{\Sigma}{N}} Tr \: (FG)(HK)  
} \end{eqnarray} 
Thus, going back to the action term (\ref{quartic}) we obtain  
\begin{eqnarray} \displaystyle{ 
Tr \: \int_0^{\frac{\Sigma}{N}} [\phi^i, \phi^j]^2 \: dx \: dy   
} \end{eqnarray} 
that is, our usual Yang-Mills terms. 
\par Let us now discuss what happens if $\theta = p/N$, $p$ and $N$ 
relatively prime. We have already discussed how this amounts to a 
``relabeling'' of the sectors of the circle by means of a permutation of 
their indices. However, it is important to point out what happens to the 
``boundary condition'' $U$ and to the ``Wilson loop'' $V$ matrices. 
\par The first point to realize is that, whatever the permutation of 
sectors may be, the ``periodic'' boundary conditions will not change. This 
is because, no matter what happens to the interactions, the free 
evolution will move along the interval in the original order; the free part 
of the lagrangian on the big circle is certainly local. 
\par What happens to the Wilson loop? If we permute the intervals 
\begin{eqnarray} \displaystyle{ 
\begin{array}{c} 
\begin{array}{lll}
1 & \longrightarrow & 1 \\
2 & \longrightarrow & (1+p)_{{\mathrm{mod}\:}N} \\ 
3 & \longrightarrow & (1+2p)_{{\mathrm{mod}\:}N} \\    
4 & \longrightarrow & (1+3p)_{{\mathrm{mod}\:}N} \\    
\end{array}        \\ 
\ldots \ldots \ldots 
\end{array} 
} \end{eqnarray} 
we will replace eq.~(\ref{c2}) with 
\begin{eqnarray} \displaystyle 
W(x)= exp \left(  i p \frac{x}{\Sigma} \right)  V^p 
\end{eqnarray} 
The non abelian part of the Wilson loop 
\begin{eqnarray} \displaystyle{ 
V= \left( 
\begin{array}{cccccc}
1 & 0 & 0 & 0 & 0 & 0 \\
0 & \xi & 0 & 0 & 0 & 0 \\
0 & 0 & \xi^2 & 0 & 0 & 0 \\
0 & 0 & 0 & \xi^3 & 0 & 0 \\
0 & 0 & 0 & 0 & \xi^4 & 0 \\
0 & 0 & 0 & 0 & 0 & \ddots \\
\end{array}
\right) 
\: \: \: \: \: \: \: \: \: \: \: \xi= e^{2 \pi i /N}
} \end{eqnarray} 
is replaced with 
\begin{eqnarray} \displaystyle{ 
V^p = \left( 
\begin{array}{cccccc}
1 & 0 & 0 & 0 & 0 & 0 \\
0 & \xi^{(1+p)_{{\mathrm{mod}\:}N} -1}  & 0 & 0 & 0 & 0 \\
0 & 0 & \xi^{(1+2p)_{{\mathrm{mod}\:}N} -1} & 0 & 0 & 0 \\
0 & 0 & 0 & \xi^{(1+3p)_{{\mathrm{mod}\:}N} -1} & 0 & 0 \\
0 & 0 & 0 & 0 & \ddots & 0 \\
0 & 0 & 0 & 0 & 0 & \ddots \\ 
\end{array}
\right) 
} \end{eqnarray} 
The abelian vector potential is now $p$ times larger, and so is the flux. 
\par Also the commutation relation is accordingly modified:  
\begin{eqnarray} \displaystyle{ 
UW= WU \, \xi^p 
} \end{eqnarray} 
(the rewriting in terms of the Wilson loop is allowed since the abelian 
part gives no contribution). 
\par To summarize, the relations in the general case are 
\begin{eqnarray} \label{annidata} \displaystyle{ 
\begin{array}{c}
\theta= \frac{p}{N} \\ \\  
\left\{ \begin{array}{l} 
\phi(x+ \frac{\Sigma}{N}, y) = U^\dagger \phi(x,y) U \\ 
\phi(x, y+ \frac{\Sigma}{N}) = \phi(y) 
\end{array} \right. \\ \\ 
W= V^p \\  \\ 
UW= WU \, \xi^p 
\end{array}
} \end{eqnarray} 
Moreover, the abelian part of the vector potential carries now $p$ units 
of magnetic flux. 

\section{Coupling constant rescaling}
\par It might be interesting to derive the behaviour of the coupling 
constant $g_{YM}$ along the process of ``folding'' the big circle into 
the small one. We know (cfr.~for example \cite{review}) that the 
Yang-Mills coupling for a 2+1 dim gauge theory describing compactification on 
a 2-torus is
\begin{eqnarray} \displaystyle{ 
g_{YM}^2 = \frac{1}{\Sigma} \frac{l_{11}^3}{L^3} 
} \end{eqnarray} 
This is the coupling constant of the non commutative geometry on the 
``big'' torus. The transition to the small torus yields a factor of $1/N$: 
\begin{eqnarray} \displaystyle{ 
\tilde{g}_{YM}^2 = \frac{1}{N \Sigma} \frac{l_{11}^3}{L^3}
} \end{eqnarray} 
Thus $g^2 N$ is kept fixed during the passage to the ``small'' torus. 

\section{Conclusions}
To summarize, we have now a prescription for the link between a super Yang 
Mills theory with large N and a noncommutative geometry with $\theta$ 
``almost irrational''. \\

\begin{itemize}
\item{Take $\theta$ and approximate it with an irreducible fraction  
$\displaystyle{\frac{p}{N}}$}
\end{itemize}
\begin{eqnarray} \nonumber \displaystyle{ 
\Updownarrow 
} \end{eqnarray} 
\begin{itemize}
\item{Build a $U(N)$ gauge theory on a torus of size 
$\displaystyle{\frac{\Sigma}{N}}$ and choose as boundary conditions those   
which are explicit in eq.~(\ref{annidata}) with $p$ units of magnetic flux.}
\end{itemize}
\vspace{0.2cm} 
Thus we find that abelian gauge theory on a noncommutative torus is equivalent 
to an appropriate limit of non abelian gauge theory on a rescaled commutative 
torus.

\end{document}